\documentclass[11pt]{article}
\usepackage[T1]{fontenc} % Use 8-bit encoding that has 256 glyphs
\usepackage[english]{babel} % Language hyphenation and typographical rules

\usepackage[left=2.5cm,top=4.5cm,right=2.5cm,bottom=3cm]{geometry} % Document margins

\usepackage{mathptmx} % Set font to Times Roman
\usepackage[scaled=.85]{helvet}

\usepackage[final]{changes}
\definechangesauthor[name={26-5},color={orange}]{26-5}
\definechangesauthor[name={27-5},color={green}]{27-5}
\definechangesauthor[name={28-5},color={red}]{28-5}

\usepackage{amsfonts}
\usepackage{amssymb}
\usepackage{amsmath}
\usepackage{amsthm}
\usepackage{graphicx}
\usepackage[justification=centering]{caption}
\usepackage{booktabs}
\usepackage{float}

\usepackage{abstract} % Allows abstract customization
\usepackage{titling} % Customizing the title section
\posttitle{\par\end{center}} % Raise the Authors up

\usepackage{titlesec} % Set section and subsection fonts
\titleformat*{\section}{\fontsize{16pt}{18pt}\selectfont \bfseries}
\titleformat*{\subsection}{\fontsize{14pt}{8pt}\selectfont \bfseries}
\titleformat*{\subsubsection}{\fontsize{12pt}{8pt}\selectfont \bfseries}

\theoremstyle{plain} % Set theorem style
\newtheorem{thm}{Theorem}

\theoremstyle{remark} % Set remark style

\usepackage{enumitem} % Customized lists
\setlist[itemize]{noitemsep} % Make itemize lists more compact

\usepackage{etoolbox} % Set Bibliography
\apptocmd{\thebibliography}{\setlength{\itemsep}{-2pt}}{}{}

\usepackage{hyperref} % For hyperlinks in the PDF

\usepackage{setspace} % Allows line spacing customization

\usepackage{fancyhdr}
\title{\fontsize{20pt}{22pt}\selectfont
{\bf The trade-off between model flexibility and accuracy of the Expected Threat model in football }}  % Article title, please make sure you are using the bold font

\author{\vspace{8pt}
Koen W. van Arem\textsuperscript{1*}, Jakob S\"ohl\textsuperscript{1}, Mirjam Bruinsma\textsuperscript{2} and Geurt Jongbloed\textsuperscript{1}\\    % Authors
\fontsize{10pt}{6pt}\selectfont    % Set font
\textsuperscript{1} Delft University of Technology, The Netherlands\\
\fontsize{10pt}{6pt}\selectfont\textsuperscript{2} AFC Ajax, The Netherlands\\
\fontsize{10pt}{6pt}\selectfont* k.w.vanarem@tudelft.nl\\   % Address
}

\date{} % Leave it empty

\begin{document}

\pagenumbering{gobble} % Omit page numbers

\maketitle

\vspace{-30pt}

\begin{abstract}
\indent With an average football (soccer) match recording over 3,000 on-ball events, effective use of this event data is essential for practitioners at football clubs to obtain meaningful insights. Models can extract more information from this data, and explainable methods can make them more accessible to practitioners. The Expected Threat model has been praised for its explainability and offers an accessible option. However, selecting the grid size is a challenging key design choice that has to be made when applying the Expected Threat model. Using a finer grid leads to a more flexible model that can better distinguish between different situations, but the accuracy of the estimates deteriorates with a more flexible model. Consequently, practitioners face challenges in balancing the trade-off between model flexibility and model accuracy.  
In this study, the Expected Threat model is analyzed from a theoretical perspective and simulations are performed based on the Markov chain of the model to examine its behavior in practice. Our theoretical results establish an upper bound on the error of the Expected Threat model for different flexibilities. Based on the simulations, a more accurate characterization of the model’s error is provided, improving over the theoretical bound. Finally, these insights are converted into a practical rule of thumb to help practitioners choose the right balance between the model flexibility and the desired accuracy of the Expected Threat model.

\end{abstract}

\section{Introduction}
An average football (soccer) match contains around 3,000 on-ball events \cite{Statsbomb2022}. This data can provide new meaningful insights for football clubs. It can, for instance, be used to study players for scouting or opponent analysis. This can help practitioners make better informed decisions. Mathematical models can be used to extract more complex information from these on-ball events \cite{ReinMemmert2016}, which can give football clubs an advantage over the competition.

However, advanced mathematical models introduced by researchers might not be adaptable or practical enough for coaches and teams \cite{HeroldGoesEA2019}. It is important to have models that can be explained in such a way that they can be understood by practitioners and utilized in practice \cite{HeroldGoesEA2019}. To achieve this, models to distill information from the in-game data should also be assessed on how explainable and interpretable they are \cite{DavisBransenEA2023}. This means that explainable models are of added value because they offer an accessible option to retrieve complex information from in-game data.

The Expected Threat \cite{Rudd2011, Singh2018} model quantifies the quality of an action, which gives detailed information about offensive player quality, and it is praised for its interpretability \cite{VanRoyRobberechts2020}. Nonetheless, practitioners still face one key design choice when applying the Expected Threat model. The Expected Threat model estimates the probability of scoring, and to do this, it divides the pitch into different in-game states. The number of in-game states, $M$, describes the model flexibility and can be chosen. A model with more game states can distinguish between more situations \cite{VanAremBruinsma2024}. On the other hand, more states decrease the accuracy of the Expected Threat model because more probabilities are estimated with the same amount of data. Increasing the amount of data is often infeasible in practice, because of the additional costs. 
This trade-off between model flexibility and accuracy hinders the application of the otherwise accessible Expected Threat model in practice. The aim of this study is to provide a rule of thumb for practitioners to manage the trade-off between model flexibility and accuracy of the Expected Threat model.

\section{Expected Threat}
The Expected Threat model \cite{Rudd2011, Singh2018} considers football as a Markov chain and is based on the idea that good actions increase the probability of scoring a goal within the possession chain. The pitch is divided into $M$ squares, and the state of the game is defined as the square where the ball-possessing player is. For each state $s$, the model then calculates the probability of scoring, denoted by $xT(s)$. The quality of an action is defined as the difference before and after the action: $\Delta xT(s_{\text{before}}, s_\text{after}) = xT(s_\text{after}) - xT(s_{\text{before}})$.

When a player has ball possession, there are two ways to score a goal: either directly score a goal or move the ball to another state with a dribble or pass and score from there. To score a goal, the player has to decide to shoot, and the player has to score the shot. If the current game state is denoted as $s$, the probability of this happening is $P(shot|s)\cdot P(goal|shot, s)$. Because $P(goal|shot, s)$ is the quantity described by Expected Goal (xG) models, this can also be denoted as $P(shot|s)\cdot xG(s)$. If the player decides not to shoot, a goal can be scored by moving the ball to each other game state $s'$ and scoring from there. The probability of scoring via the game state $s'$ can be written as $T_{s\to s'}\cdot xT(s')$, where $T_{s\to s'}$ is the probability of transitioning from $s$ to $s'$. This means that the probability of scoring from game state $s$ is
\begin{equation}\label{eq: expected threat defining equation}
    xT(s) = P(shot|s)\cdot xG(s) + \sum_{s'}T_{s\to s'}\cdot xT(s').
\end{equation}

In practice, the values of $P(shot|s)$, $xG(s)$, and $T_{s\to s'}$ are estimated by counting the occurrences of these events in the data set. When these are estimated, the only unknowns in \eqref{eq: expected threat defining equation} are the values $xT(s)$. Because \eqref{eq: expected threat defining equation} holds for each state $s$, it gives a system of equations, which is generally solved using an iterative algorithm. In this way, the probability of scoring from state $s$ is estimated.

Due to randomness in the training data, errors are made in the estimation of $P(shot|s)$, $xG(s)$, and $T_{s\to s'}$. These estimation errors cause errors in the estimated $xT$-values. For practitioners, it is important to have a bound on these errors. The model error in this study is defined as the maximal difference between the true and the estimated xT-values, denoted by $||xT-\widehat{xT}||_\infty$. The distribution of this error depends on the number of training points $N$ and the number of game states $M$ and can be used to describe the trade-off between model flexibility, described by $M$, and model accuracy.

Using the properties of the Expected Threat model, it is possible to derive probabilistic bounds on the error of the model. For this purpose, the transition matrix is the matrix with the transition probabilities $T_{s\to s'}$ as entries. A summary of our theoretical results is described in the following theorem.

\begin{thm}
    Let $g\in \mathbb{R}^M$ be defined by $g_s=P(shot|s)\cdot xG(s)$ and let $T$ be the transition matrix. Assume that $||T||_\infty<1$ and that for the estimated transition matrix $||\hat{T}||_\infty<1$. Moreover, assume that the estimates of the quantities in the Expected Threat model are obtained by taking averages of $N$ independent Bernoulli random variables. Then the following bounds hold with probability at least $1-\alpha$:
    \begin{equation}
    ||\widehat{xT} - xT||_\infty \leq \frac{1}{1-||T||_\infty}\left(M\sqrt{\frac{\log(2M^2/\alpha)}{2N}}+\sqrt{\frac{\log(2M/\alpha)}{2N}}\right)\leq \frac{2}{1-||T||_\infty}\left(M\sqrt{\frac{\log(2M^2/\alpha)}{2N}}\right).
\end{equation}
More specifically, the term $M\sqrt{\frac{\log(2M^2/\alpha)}{2N}}$ corresponds to the error in estimating $T$ and  $\sqrt{\frac{\log(2M/\alpha)}{2N}}$ to the error in estimating $g$.
\end{thm}
This theorem shows that the error in estimating the xT-values is of the order $O(M\sqrt{\log(M)}/\sqrt{N})$. However, the results also suggest that, in practice, implemented models might fall under a finite sample regime where the error in g is still larger, and where a faster decay of the error can be observed.

\section{Methods}
To find the error distribution of the  Expected Threat model in practice, simulations were performed based on the Markov chain underlying the model.
The data used for this simulation study is obtained from the openly available Statsbomb data set \cite{StatsBombOpenData}. All available games from the Premier League, Ligue 1, Serie A, La Liga, and the Bundesliga were used. The events that did not describe passes, dribbles, errors, clearances, or shots were filtered out. This resulted in a data set of approximately 4,000,000 events, which is equivalent to around 6.5 full seasons of one league.

In this research, the maximal model error was studied for Expected Threat models with different discretization grids, and thus for different flexibilities $M$, which are described in \autoref{tab:grids and sample sizes}. For each grid size $M$, one Expected Threat model was calculated with the Statsbomb data. These models were assumed to be the ground truth for elite male leagues within the scope of this study. Because the data set is relatively large, this is a reasonable assumption.

\begin{figure}[h]
    \centering
    % \begin{minipage}[t]{0.38\textwidth}
    \begin{minipage}[t]{0.69\textwidth}
        \vspace{0pt}
        \begin{table}[H] 
            \centering
            \begin{tabular}{c c c || c}
                \toprule
                \( n_x \) & \( n_y \) & \( M = n_x \cdot n_y \) & N \\
                \midrule
                16  & 12  & 192  & 100,000  \\
                32  & 24  & 768  & 130,000  \\
                40  & 30  & 1200 & 170,000  \\
                48  & 36  & 1728 & 240,000  \\
                56  & 42  & 2352 & 370,000  \\
                64  & 48  & 3072 & 630,000  \\
                    &     &      & 1,300,000\\
                    &     &      & 4,000,000\\
                \bottomrule
            \end{tabular}
            \caption{Grid configurations with corresponding \(M\) values and the sample sizes \(N\) of the simulations.}
            \label{tab:grids and sample sizes}
        \end{table}
    \end{minipage}
    \hfill
\end{figure}

Each of these ground truth models describes a Markov chain, which can be used to resample a new data set. For each ground truth model, new data sets of different sizes $N$ were resampled, which were then used to train resampled Expected Threat models. The model error was then obtained by calculating the maximal absolute difference between the ground truth $xT$-values and the $xT$-values based on the resampled data. This process was repeated 1,000 times for each combination of $M$ and $N$ as described in \autoref{tab:grids and sample sizes}. This created 48,000 data points describing the model error, the grid size $M$ and the number of data points $N$.

The goal of this research is to obtain more insight into the trade-off between the model flexibility governed by $M$ and the model accuracy, described by the model error. Because situations with $M\log(M)/\sqrt{N}\geq 15$ resulted in errors too large for practice, these were filtered out. This makes it possible to study the distribution of the errors in this setting, which is interesting for practitioners

Using the simulated data, the distribution of the maximal model error could be studied. To do this, the following lognormal model was assumed to describe the maximal model error:
\begin{equation}\label{eq: assumed lognormal model}
    ||\widehat{xT} - xT||_\infty = C\frac{M^\alpha}{(\sqrt{N})^\beta}e^\varepsilon, \quad \text{ where } \varepsilon\sim N(0,\sigma^2).
\end{equation}
The model error, which is a maximal absolute difference, is known to be positive. Additionally, the theoretical results indicated that both the mean and spread of the error are small if $M$ is small and $N$ is large. This makes the lognormal model a reasonable assumption. Moreover, this model describes the powers of the variables $M$ and $N$, which are unknown because the theoretical results only gave an upper bound.

If $c=\log(C)$, \eqref{eq: assumed lognormal model} is equivalent with
\begin{equation}\label{eq: assumed lognormal ols model}
    \log(||\widehat{xT} - xT||_\infty) = c + \alpha\log(M) - \beta \log(\sqrt{N}) + \varepsilon, \quad \text{ where } \varepsilon\sim N(0,\sigma^2).
\end{equation}
This formulation gives a linear model with normal residuals. Therefore, ordinary least squares (OLS) can be applied to estimate $c,\alpha,\beta$, and $\sigma^2$.

\section{Results}

\begin{figure}[htb]
    \centering
    \begin{minipage}[t]{0.55\textwidth}
        \vspace{0pt}

        \begin{table}[H]
            \centering
            \resizebox{\linewidth}{!}{%
            \begin{tabular}{lclc}
                \toprule
                \textbf{Dep. Variable:}    & $||\widehat{xT} - xT||_\infty$ & \textbf{  R-squared:         } &     0.835  \\
                \textbf{Model:}            &         OLS          & \textbf{  Adj. R-squared:    } &     0.835  \\
                \textbf{No. Observations:} &         23000 & \textbf{  Log-Likelihood:    } &   -9297.8  \\
                \textbf{Df Residuals:}     &         22997        & \textbf{  AIC:               } & 1.860e+04  \\
                \textbf{Df Model:}         &             2        & \textbf{  BIC:               } & 1.863e+04  \\
                \bottomrule
            \end{tabular}}
            % \vspace{0.5em}
    
            \resizebox{\linewidth}{!}{%
            \begin{tabular}{lcccccc}
                \toprule
                                        & \textbf{coef} & \textbf{std err} & \textbf{t} & \textbf{P$> |$t$|$} & \textbf{[0.025} & \textbf{0.975]}  \\
                \midrule
                \textbf{c}              & -1.8758 & 0.026 & -72.138 & 0.000 & -1.927 & -1.825 \\
                $\mathbf{\alpha}$       & 0.9898  & 0.003 & 330.569 & 0.000 & 0.984  & 0.996  \\
                $\beta$                 & 1.0416  & 0.004 & 238.837 & 0.000 & 1.033  & 1.050  \\
                \bottomrule
                \textbf{Variance residuals} & 0.1314 \\
            \end{tabular}}
            \captionof{table}{Summary of the OLS model fitted on the log maximal model error for the data points with $M\log(M)/\sqrt{N} < 15$.}
            \label{tab:OLS summary lognormal distribution}
        \end{table}
        
    \end{minipage}
    \hfill
    \begin{minipage}[t]{0.4\textwidth}
        \vspace{0pt}
        \centering
        \includegraphics[width=\linewidth]{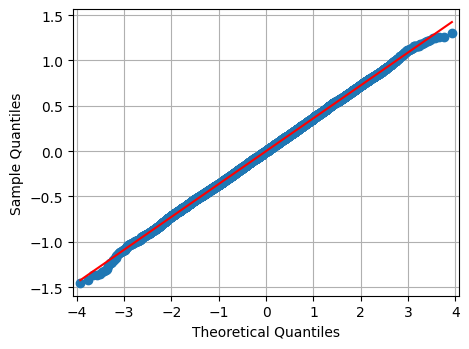}
        \caption{A QQ plot of the residuals of the fitted OLS model.}
        \label{fig:qq plot residuals}
    \end{minipage}
\end{figure}

The summary of the ordinary least squares applied to \eqref{eq: assumed lognormal ols model} is given in \autoref{tab:OLS summary lognormal distribution}. It shows that the adjusted $R^2$ is 0.835, which indicates that the model is able to explain most variance of the errors with $M$ and $N$. Additionally, \autoref{fig:qq plot residuals} shows the QQ plot of the residuals of the lognormal OLS model. It is visible that the residuals indeed seem to be well-described by the lognormal OLS model. Thus, it can be concluded that the model in \eqref{eq: assumed lognormal ols model} provides a good description of the distribution of the maximal error of the Expected Threat model. With the found values for $\alpha, \beta$ and $\sigma^2$, the distribution of the model can be described by
\begin{equation}\label{eq: fitted distribution lognormal}
    ||\widehat{xT} - xT||_\infty = e^{-1.8758}\frac{M^{0.9898}}{(\sqrt{N})^{1.0416}}e^\varepsilon, \quad \text{ where } \varepsilon\sim N(0,0.1314).
\end{equation}

The values found for $\alpha$ and $\beta$ are close to 1, although significantly different according to the confidence intervals in \autoref{tab:OLS summary lognormal distribution}. This means that the maximal error of the model is of order $O(M^{0.9898}/(\sqrt{N})^{1.0416})$, which indicates that the error, in practice, is of a lower order than established by the theoretical results.

\section{Rule of thumb}
\begin{figure}[ht]
    \centering
    \begin{minipage}[t]{0.48\textwidth}
        \vspace{0pt}
        \centering
        \includegraphics[width=\linewidth]{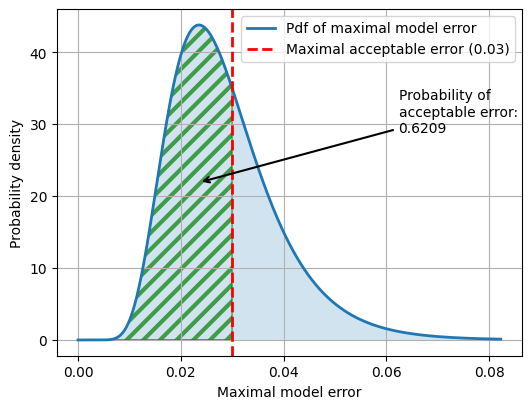}
        \caption{The distribution of the maximal model error of the Expected Threat model in \cite{Singh2018}, where $M=192$ and $N=620{,}000$.}
        \label{fig:singh error distribution}
    \end{minipage}
    \hfill
    \begin{minipage}[t]{0.48\textwidth}
        \vspace{0pt}
        \centering
        \includegraphics[width=\linewidth]{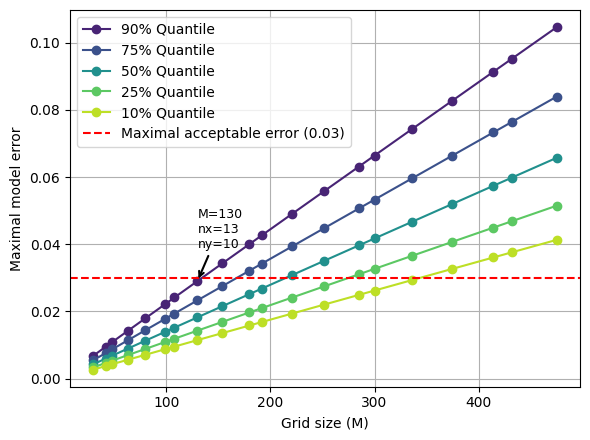}
        \caption{The quantiles of the maximal model error of an Expected Threat model with $N=2{,}480{,}000$ training data points for different grid sizes $M$.}
        \label{fig: rule of thumb}
    \end{minipage}
\end{figure}

With the distribution of the model error, it is possible to give guidance to practitioners on how to use the Expected Threat model. If they have an existing model, the distribution of the error in \eqref{eq: fitted distribution lognormal} can be used to describe the distribution of the error of their model. For example, consider the Expected Threat model by Singh \cite{Singh2018}, which has a $16\times12$ grid and one season of the Premier League as training data. This corresponds to $M=16\cdot12 = 192$ game states and roughly $N=620,000$ training data points. In the experience of consulted experts, errors smaller than 0.03 are acceptable for scouting purposes. The distribution of the maximal error of this model is visualized in \autoref{fig:singh error distribution}. It indicates that there is a 62.09\% chance of having an error that is lower than 0.03. This means that there is a reasonable chance of this model having an acceptable error.

When training a new model, the number of game states $M$ has to be chosen. It is desirable to have a low model error with a high probability. On the other hand, a more flexible model can better distinguish between in-game situations. To balance this trade-off, a reasonable idea would be to choose the most flexible model with an acceptable statistical error. Through consultation with experts, it was established that the Expected Threat model is sufficiently reliable for scouting purposes when the maximal model error is smaller than 0.03 with a 90\% probability. This can be reformulated as the following rule of thumb: \emph{select the most flexible model (highest $M$) such that the maximal model error is smaller than 0.03 with a 90\% probability.}

To illustrate this rule of thumb, suppose a practitioner wants to train an Expected Threat model on a data set of 2,480,000 data points. This corresponds to data of 4 league seasons. \autoref{fig: rule of thumb} shows different quantiles of the error for values of $M$ for this number of data points $N$ based on \eqref{eq: fitted distribution lognormal}. The results show that the maximal $M$ with a 90\% quantile smaller than 0.03 is $M=130$, which corresponds to a $13\times10$ grid. This means that the rule of thumb gives that a $13\times10$ grid yields the most flexible model with an acceptable model error. In this way, the rule of thumb provides guidance to practitioners on how to deal with the trade-off between model flexibility and accuracy, which makes the accessible Expected Threat model even more easily applicable for practitioners.

\noindent
\begin{minipage}[t]{0.45\textwidth}
    \vspace{0pt}
    \section*{Acknowledgements}
    The authors would like to thank StatsBomb for making the StatsBomb Open Data publicly available.
\end{minipage}%
\hfill
\begin{minipage}[t]{0.45\textwidth}
    \vspace{25pt}
    \centering
    \includegraphics[width=\linewidth]{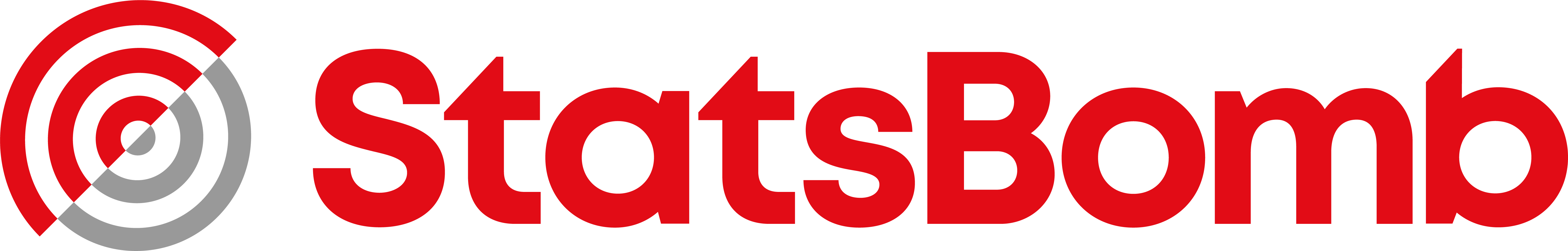}
\end{minipage}

\appendix

\end{document}